\documentclass[aps,pra,reprint,superscriptaddress]{revtex4-2}

\usepackage{physics,amsfonts,amsmath,graphicx,dcolumn,bm,array,color,hyperref,amssymb,setspace,empheq,braket,multirow}
\hypersetup{
  colorlinks   = true,
  urlcolor     = blue,
  linkcolor    = blue,    
  citecolor    = blue
}
\begin{document}

\title{Temporal quantum eraser: Fusion gates with distinguishable photons}

\author{Ziv Aqua}
\email{zivaqua@gmail.com}
\affiliation{AMOS and Chemical and Biological Physics Department, Weizmann Institute of Science, 76100 Rehovot, Israel}
\author{Barak Dayan}
\affiliation{AMOS and Chemical and Biological Physics Department, Weizmann Institute of Science, 76100 Rehovot, Israel}
\affiliation{Quantum Source Labs, Israel}

\date{\today}
             
\begin{abstract}
Linear-optics gates, the enabling tool of photonic quantum information processing, depend on indistinguishable photons, as they harness quantum interference to achieve nonlinear operations. Traditionally, meeting this criterion involves generating pure identical photons, a task that remains a significant challenge in the field. Yet, the required indistinguishability is linked to the spatial exchange symmetry of the multiphoton wavefunction and does not strictly necessitate identical photons. Here, we show that the ideal operation of two-photon gates, particularly fusion gates, can be recovered from distinguishable photons by ensuring the exchange symmetry of the input photonic state. To this end, we introduce a temporal quantum eraser between a pair of modally-impure single-photon sources, which heralds the symmetry of the generated two-photon state. We demonstrate this mechanism in two relevant platforms: parametric photon pair generation and single-photon extraction by a single quantum emitter. The ability to lift the requirement for identical photons bears considerable potential in linear-optics quantum information processing.
\end{abstract}
\maketitle

\section{Introduction}
Photonic quantum information with linear optics utilizes quantum interference in photon detection events to achieve nonlinear operations. Specific detection events at designated output ports of the linear system herald the success of the desired operation. Notable examples include Bell state measurement used for teleportation and for entanglement swapping \cite{bennett1993teleporting,zukowski1993event,braunstein1995measurement,pan1998exp}, construction of GHZ states \cite{varnava2008good,gubarev2020improved}, and controlled-NOT gate for linear-optics quantum computing \cite{knill2001scheme,o2003demonstration,carolan2015universal}. Of particular interest are fusion gates, first introduced as an efficient method for generating photonic cluster states \cite{browne2005resource} utilized as a resource for measurement-based quantum computation \cite{raussendorf2003measurement}. More recently, these gates are at the heart of fusion-based quantum computation \cite{bartolucci2023fusion}, serving as destructive entangling measurements and as a tool for constructing the required small entangled resource states \cite{bartolucci2021creation}.

\begin{figure}
    \centering
    \includegraphics[width=6cm]{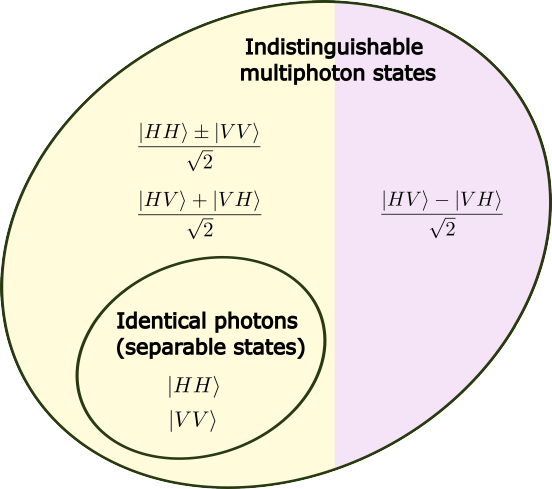}
    \caption{Useful multiphoton states for linear-optics quantum information processing. The cases in which all the input photons are identical (smaller ellipse), i.e occupying identical modes that propagate in different rails, is only a subset of the useful input states. The broader set comprises of multiphoton states that may involve distinct modes varying in certain degrees of freedom, yet exhibit a defined symmetry with respect to exchange of the input rails. While we chose to present states in the discrete polarization degree of freedom as examples, our discussion is more general, encompassing multiphoton states with continuous degrees of freedom as well, such as spectro-temporal modes. The yellow area represents states that are perfectly symmetric to exchange between the input modes (i.e. symmetric to switching the order of the letters in the ket), such as identical photons and symmetric Bell states. The violet area corresponds to states with other exchange symmetries, e.g. the antisymmetric singlet state.} 
    \label{fig:venn}
\end{figure}

Quantum interference arises from the contribution of indistinguishable paths to the final state, making indistinguishability imperative to the performance of linear-optics quantum information processing \cite{kok2007linear,flamini2018photonic}. This requirement has promoted the use of identical single photons, which, apart from propagating in distinct rails, share identical properties across all degrees of freedom - polarization, tempo-spectral and spatial modes. Generating such identical photons from different sources presents one of the primary challenges in the field, often involving significant efforts in terms of complexity and resources \cite{sparrow2018quantum,marshall2022distillation,meyer2017limits,faurby2024purifying}.

However, indistinguishable paths do not necessarily require identical photons; it is sufficient that there is no distinguishing information indicating which input mode resulted in which detection event \cite{kwiat1992observation,kim2005quantum,jones2020multiparticle}. In fact, quantum interference is completely dictated by the symmetry of the multiphoton wavefunction with respect to exchange between the inputs \cite{ou2006temporal,tillmann2015generalized}. In other words, as illustrated in Figure \ref{fig:venn}, linear-optics quantum information processing can utilize multiphoton wavefunctions with a defined exchange symmetry. Identical photons, defined by a separable symmetric wavefunction, represent a special case within a broader class of indistinguishable multiphoton states. It should be emphasized that indistinguishable wavefunctions, apart from identical photons, must exhibit some degree of entanglement. 

\begin{figure}
    \centering
    \includegraphics[width=0.95\linewidth]{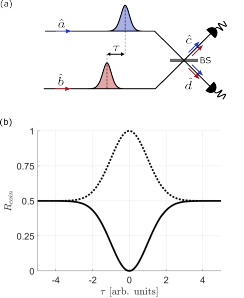}
    \caption{The HOM experiment. (a) Two photons are directed towards the input ports of a BS. A relative delay between the incoming modes, denoted by $\tau$, is introduced to adjust the overlap between their wavepackets. The HOM effect is observed by measuring the probability of two detection events at different output ports, referred to as the coincidence rate, $R_{coin}$. (b) For a spatially symmetric (antisymmetric) two-photon wavefunction, the paths leading to photon antibunching (bunching) at the output ports destructively interfere, resulting in a coincidence rate of zero (one) at zero relative delay. When $\tau$ is set such that the overlap between the modes vanishes, the paths become completely distinguishable and the photons have an equal probability of leaving in either output port, yielding a coincidence rate of half. This leads to the characteristic HOM dip (peak) for a symmetric (antisymmetric) two-photon state, as indicated by the solid (dotted) line.}
    \label{fig:HOM}
\end{figure}

Consider the Hong-Ou-Mandel (HOM) effect, a fundamental two-photon interference phenomenon \cite{hong1987measurement}, which has long served as a standard for characterizing photon indistinguishability. This effect involves interfering two photons on a 50:50 beam splitter (BS), as depicted in Figure \ref{fig:HOM}(a). When the two photons are identical, the paths leading to a single photon at each of the output ports of the BS interfere destructively. This results in photon bunching, where the two photons exit the BS together through either output port. Yet, it can be easily confirmed using the BS input-output relations that the HOM effect is truly a measure of the spatial symmetry of the two-photon state under exchange; a symmetric (antisymmteric) state leads to perfect bunching (antibunching) at the output ports. Therefore, any spatially symmetric or antisymmetric two-photon state, even of distinct internal degrees of freedom, would yield perfect two-photon interference and can hence be utilized reliably in two-photon linear-optics gates. 

Note that due to the bosonic nature of photons, antisymmetry in the spatial degree of freedom must be paired with antisymmetry in another degree of freedom. These aspects of two-photon interference have been discussed \cite{wang2006quantum,fedrizzi2009anti} and demonstrated with polarization Bell pairs \cite{walborn2003multimode,dayan2007spectral}, high-dimensional orbital angular momentum states \cite{zhang2016engineering} and hyperentangled photons \cite{liu2022hong}. A common feature of these demonstrations is that the photons are produced probabilistically in an unheralded manner, thereby limiting their applicability in protocols necessitating on-demand single-photon sources (SPS).

Photon distinguishability, i.e. asymmetry of the photonic state with respect to exchange, can stem from a physical distinction between the modes of the different sources, e.g. different phase matching conditions in non-linear media, or different resonant frequencies of quantum emitters. Despite such distinctions, quantum interference can still be recovered and harnessed for entanglement generation, for instance, through time-resolved measurements and active feedforward \cite{zhao2014entangling,wang2018experimental,yard2024on}. Yet, even when considerable efforts are made to ensure the sources are nearly identical, photon distinguishability may still arise from the generation process itself, where information regarding the mode of the output photon can leak out \cite{alexander2024manufacturable}. When this information is not actively taken into account, the source suffers from mode impurity, resulting in a mixed state for the generated photons, and accordingly, they are at least partially distinguishable.
 
In this work, we focus on the case where identical copies of such modally-impure sources are used. In particular, we examine sources in which temporal information associated with the output mode of the photon is carried by additional photons emitted from the same source. We theoretically demonstrate that by applying a temporal quantum eraser (TQE) on the information from a pair of sources, we can project the two generated photons into a purely symmetric or antisymmetric state. This results in perfect retrieval of the two-photon interference in otherwise distinguishable photons, and enables the reliable performance of two-photon linear optics gates, e.g. fusion gates. 

The remainder of the paper is structured as follows. First, we describe the TQE mechanism in two relevant examples. In Sec. \ref{sec:NL}, we present its application in the commonly used probabilistic SPS based on parametric photon pair generation in non-linear (NL) medium. Then, in Sec. \ref{sec:Lambda}, we consider a deterministic SPS based on single-photon extraction by a single quantum emitter, where the temporal impurity is inherent, making the use of the TQE even more vital. Intriguingly, we also show that the TQE configuration in this case can also be harnessed for the deterministic generation of cat states. Subsequently, in Sec. \ref{sec:Fusion}, we present the proper interpretation of detection events in fusion gates employing purely symmetric or antisymmetric two-photon states. Finally, in Sec. \ref{sec:Diss}, we offer concluding remarks, discussing the advantages of this approach and its limitations.

\section{Parametric photon pair generation}
\label{sec:NL}
Photon pairs can be stochastically generated in NL optical media through spontaneous parametric down-conversion and spontaneous four-wave mixing, involving materials with $\chi^{(2)}$ and $\chi^{(3)}$ electric susceptibilities, respectively \cite{burnham1970observation,pittman2002single,fiorentino2002all,rarity2005photonic}. In these processes, incident pump photons are converted into photon pairs, where each pair consists of a signal photon and an idler photon, emitted into separate modes. Ideally, detection of an idler photon in one mode, ensures the presence of the signal photon in the other, thus providing a reliable heralded SPS. Yet, in practical systems, when multiple pairs are produced within the same pump pulse, photon loss and detector inefficiency introduce number impurity to the SPS. To minimize this effect, the generation process is adjusted such that the mean number of photon pairs generated per pulse is significantly lower than one. In turn, the low generation probability can be mitigated by multiplexing several sources, potentially enabling on-demand SPS \cite{ma2011experimental,collins2013integrated,xiong2016active}.

In order to keep our analysis simple, we examine photons pairs emitted into two guided modes (e.g. two different single-mode waveguides, or opposite directions of the same waveguide), acknowledging that the outcomes are applicable to any choice of separate modes. In what follows, we neglect terms involving multiple pairs. Under these assumptions, the state of the signal and idler photons is given by,
 \begin{equation}
     \ket{\psi^{(1)}_\text{NL}} = \int dt_1 dt_2 \Phi(t_1,t_2) a_i^\dagger(t_1)a_s^\dagger(t_2) \ket{0}
 \end{equation}
where $\Phi(t_1,t_2)$ is determined by the phase-matching condition in the NL medium and the bandwidth of the pump. Energy conservation in the process leads to time-energy entanglement between the generated photons. Therefore, temporal information on the signal can be inferred from the detection time of the idler, thus introducing a degree of distinguishability, even when the sources are identical. 
This temporal entanglement is one of the main factors that currently limit the HOM visibility between independent sources \cite{mosley2008heralded,kaltenbaek2009high,tanida2012highly,xiong2016active,chen2019indistinguishable,llewellyn2020chip,paesani2020near,pickston2021optimised,alexander2024manufacturable}.

\begin{figure}[t]
    \centering    
    \includegraphics[width=0.95\linewidth]{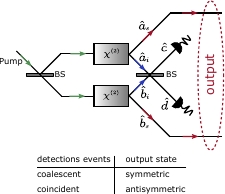}
    \caption{Schematic description of a TQE between two NL sources. A pump pulse is sent to two $\chi^{(2)}$ media, stochastically generating pairs of signal and idler photons. The idler modes, $\hat{a}_i$ and $\hat{b}_i$, are mixed on a BS and then directed to detectors. Two detection events within the same pulse ensure the presence of two signal photons. The detection pattern is used to herald the symmetry of the two-photon wavefunction; coalescent (coincident) detection events in $\hat{c}$ and $\hat{d}$ herald a symmetric (antisymmetric) two-photon state at the output.}
    \label{fig:NL}
\end{figure}

Nevertheless, the temporal information carried by idler photons from two identical sources can be 'erased' by mixing them on a BS, effectively implementing a HOM experiment between the idler photons, as depicted in Figure \ref{fig:NL}. Consider the state of two photon pairs generated from NL sources with the same pump laser and phase-matching condition,
\begin{alignat}{1}
     \ket{\psi^{(2)}_\text{NL}} = &\int dt_1 dt_2 dt_3 dt_4 \Phi(t_1,t_2) \Phi(t_3,t_4) \\       &\times a_i^\dagger(t_1)a_s^\dagger(t_2)b_i^\dagger(t_3)b_s^\dagger(t_4) \ket{0} \nonumber
 \end{alignat}
where $a_s$ and $a_i$ ($b_s$ and $b_i$) represent the signal and idler of the first (second) source. Utilizing the the fact that the sources are identical, we can rewrite the state as,
\begin{alignat}{2}
    \ket{\psi^{(2)}_\text{NL}} = & \frac{1}{2}\int dt_1&& dt_2 dt_3 dt_4 \Phi(t_1,t_2) \Phi(t_3,t_4) \label{eq:NL} \\
        & \times \sum_{m=0}^1 &&\Bigl(a_i^\dagger(t_1)b_i^\dagger(t_3) + (-1)^m a_i^\dagger(t_3)b_i^\dagger(t_1)\Bigr) \nonumber \\
      & && \times \Bigl(a_s^\dagger(t_2)b_s^\dagger(t_4) +(-1)^m a_s^\dagger(t_4)b_s^\dagger(t_2)\Bigr)\ket{0} \nonumber
\end{alignat}

As indicated by Eq. \ref{eq:NL}, the symmetry of the idlers is perfectly correlated with that of the signals, where the $m\!=\!0$ and $m\!=\!1$ terms represent the symmetric and antisymmetric states, respectively. Consequently, since the symmetry of a two-photon state dictates the outcome of a HOM experiment, by interfering the idlers on a BS we can herald the symmetry of the signal photons. 

Applying the following BS transformation,
\begin{align}
    c = \frac{a_i+b_i}{\sqrt{2}} \;\;\; ; \;\;\; d = \frac{a_i-b_i}{\sqrt{2}} \nonumber
\end{align}
to the idler modes in Eq. \ref{eq:NL} results in,
\begin{alignat}{2}
    \ket{\psi^{(2)}_\text{NL}} = &\frac{1}{2}\int &&dt_1 dt_2 dt_3 dt_4 \Phi(t_1,t_2) \Phi(t_3,t_4) \label{eq:NLf}\\
        & \times \Biggl[&&\Bigl(c^\dagger(t_1)c^\dagger(t_3) - d^\dagger(t_1)d^\dagger(t_3)\Bigr) \nonumber \\ 
      & && \begin{aligned}[t]
            &\times \Bigl(a_s^\dagger(t_2)b_s^\dagger(t_4) + a_s^\dagger(t_4)b_s^\dagger(t_2)\Bigr) \nonumber \\
            &- \Bigl(c^\dagger(t_1)d^\dagger(t_3) - c^\dagger(t_3)d^\dagger(t_1)\Bigr) \nonumber\\
            &\times \Bigl(a_s^\dagger(t_2)b_s^\dagger(t_4) - a_s^\dagger(t_4)b_s^\dagger(t_2)\Bigr)\Biggr]\ket{0} \nonumber
        \end{aligned}
\end{alignat}

As evident from Eq. \ref{eq:NLf}, coalescent or coincident detection events in modes $\hat{c}$ and $\hat{d}$ project the signal photons to a purely symmetric or antisymmetric state, respectively. Therefore, applying the TQE to the idler photons from two NL sources enables us to herald the symmetry of the signal photons, thus retrieving perfect two-photon interference from otherwise partially distinguishable photons. However, two detection events in $\hat{c}$ and $\hat{d}$, can also occur if one of the sources produces two pairs of signal and idler photons. This leads to a two-photon state with a defined symmetry, where both signal photons occupy a single waveguide (see Appendix \ref{app:NL_multipair}). Such a scenario limits the use of this scheme in protocols where qubits are encoded in single photons. Furthermore, to achieve an on-demand SPS, heralding on pairs of idler photons would require multiplexing of more sources, as it reduces the average number of generated signal photons compared to heralding on a single idler per source. Nonetheless, the ability to tailor the temporal symmetry of a two-photon state may serve as a valuable resource in its own right. In the following section, we introduce a deterministic SPS that avoids these limitations when combined with a TQE.

\section{Single-photon extraction}
\label{sec:Lambda}

 \begin{figure*}
    \centering
    \includegraphics[width=0.95\linewidth]{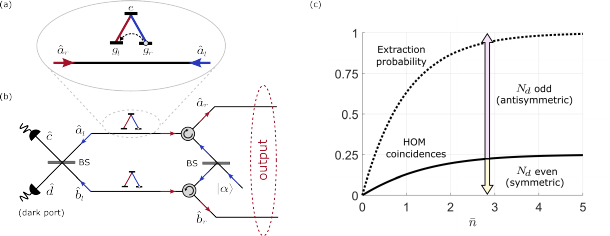}
    \caption{$\Lambda$-type quantum emitter source. (a) The basic principle of SPRINT and single-photon extraction. Two orthogonal modes $\hat{a_r}$ and $\hat{a_l}$ couple exclusively to transitions $\ket{g_l}\! \leftrightarrow\! \ket{e}$ and $\ket{g_r}\! \leftrightarrow\! \ket{e}$, respectively. A quantum emitter prepared in $\ket{g_r}$ will deterministically reflect (extract) the first photon arriving from a pulse in mode $\hat{a_l}$, and undergo Raman transition to the other ground state, thereby becoming transparent to any subsequent photons \cite{rosenblum2016extraction}. (b) Schematic description of a TQE between two $\Lambda$-type quantum emitter sources. An incoming coherent pulse, $\ket{\alpha}$, impinges on a balanced MZI, where each arm is coupled to a $\Lambda$-system prepared in its left ground state. Circulators divert the extracted single photons in both arms to the different output ports. (c) The performance of a SPS based on extraction of a single photon from a classical coherent pulse. Success probability of extracting exactly one photon (dotted), and HOM coincidence rate with two such sources (solid) as a function of the average number of photons in the incident pulse, $\bar{n}$. As evident, the mode impurity of the generated photons leads to non-zero HOM coincidence rate. However, when incorporating a TQE between a pair of such sources, the parity of the number of detected photons $N_d$ in the dark port of the interferometer (which is not dark between the two extraction events) indicates the exchange symmetry of the two-photon state, yielding perfect HOM interference with either zero or unit coincidence rate.} 
    \label{fig:Lambda}
\end{figure*}

Single photon Raman interaction (SPRINT) \cite{shomroni2014all,rosenblum2016extraction,rosenblum2017analysis} occurs in a $\Lambda$-type three-level quantum emitter, where the two transitions are exclusively coupled to two orthogonal photonic modes, e.g. two directions of a waveguide (see Figure \ref{fig:Lambda}(a)). In this process, an incident single photon interacts with a $\Lambda$-system initialized in the bright state associated with the incoming mode. Destructive interference in that mode 'forces' the quantum emitter to toggle to the corresponding dark state and reflect a photon in the orthogonal mode. As proposed in \cite{gea2013photon} and demonstrated in \cite{rosenblum2016extraction}, when sending a multiple-photon pulse to a quantum emitter in the configuration that leads to SPRINT, the first photon is reflected and the quantum emitter is pushed to its corresponding dark state, thereby transmitting the subsequent photons of the pulse. In this way, a single photon can be extracted from a classical coherent pulse with near-unity success probability. Ideally, the only failure mechanism arises from the absence of a photon in the incident coherent pulse, which decreases exponentially with the average number of photons in that pulse, denoted by $\bar{n}$ (see Figure \ref{fig:Lambda}(c)). Upon an incoming coherent state pulse $f(t)$ in mode $\hat{a}_l$ and a quantum emitter prepared in $\ket{g_r}$, the final state of the photonic field associated with the emitter ending in the respective dark state, $\ket{g_l}$, is given by \cite{gea2013photon},
\begin{alignat}{1}
    \ket{\psi^{(1)}_\Lambda} \propto &\int_{-\infty}^{\infty} dt f(t) a_r^\dagger(t) \label{eq:extract} \\
    &\times \text{exp}\biggl(\sqrt{\bar{n}} \int_{t}^{\infty} dt' f(t') a_l^\dagger(t') \biggr) \ket{0} \nonumber
\end{alignat}

As evident from the limits of integration in Eq. \ref{eq:extract}, the transmitted photons carry information about the temporal mode of the extracted single photon. Since the reflected single-photon pulse is guaranteed to end before the quantum emitter becomes transparent to subsequent photons, a detection event in transmission induces a sharp cut-off to the extracted pulse. When tracing over the transmitted field, this time-entanglement leads to a single photon in a mixed state of temporal modes. Therefore, the mode purity of extracted photons decreases with $\bar{n}$, and the HOM coincidence rate between a pair of such photons approaches $1/4$ as $\bar{n}\!\rightarrow\! \infty$  \cite{gorshkov2013dissipative} (see Figure \ref{fig:Lambda}(c)).

Incorporating a TQE between two extraction sources allows the temporal information carried by the transmitted fields to be utilized for heralding the symmetry of the two extracted photons, thereby recovering perfect two-photon interference. Consider the setup depicted in Figure \ref{fig:Lambda}(b), in which modes $\hat{a}_l$ and $\hat{b}_l$ are part of a balanced Mach-Zehnder interferometer (MZI). The relative phase between the two arms of the interferometer is set such that, in the absence of quantum emitters, a coherent state at the input interferes constructively (destructively) on $\hat{c}$ ($\hat{d}$), referred to as the bright (dark) port. The extraction procedure begins with the two emitters in their right ground state and an incident coherent pulse with an average photon number $\bar{n}$. We assume that $\bar{n}$ is sufficiently large such that the vacuum component in the two arms of the interferometer is negligible and both photon extractions occur with a unit success probability. Initially, the emitters in their reflective state do not allow the field to be transmitted. Once a single photon has been extracted by only one of the quantum emitters, it becomes transparent and the transmitted field impinging on the output BS can lead to detection events in both the dark and bright port. When the second photon has been extracted and both emitters are transparent, the transmitted fields interfere such that photons are detected only on the bright port. 

By adopting Eq. \ref{eq:extract} for both arms of the MZI, applying the appropriate BS transformation and rearranging the terms, we obtain the final state of the field associated with both quantum emitters in their respective dark states (see Appendix \ref{app:SPRINT_TQE} for a detailed derivation),
\begin{alignat}{2}
    \ket{\psi^{(2)}_\Lambda}  \propto &\int_{-\infty}^\infty &&dt \int_t^\infty dt' f(t) f(t')  \label{eq:Lambda_TQE} \\
    &\times \text{exp}&&\biggl( \sqrt{\bar{n}}  \int_{t'}^\infty d\tau f(\tau) c^\dagger(\tau) \biggl) e^{Z_c}  \nonumber \\
    &\times \sum_{m=0}^1 &&\Bigl(a_r^\dagger(t)b_r^\dagger(t') +  (-1)^m a_r^\dagger(t')b_r^\dagger(t) \Bigr) \nonumber \\ 
    & && \times \Bigl(e^{Z_d} + (-1)^m e^{-Z_d}\Bigr) \ket{0} \nonumber
\end{alignat}
where we define the creation operators for the bright and dark port, acting between the first extraction event, at time $t$, and the second extraction event, at time $t'$,
\begin{align}
    Z_\xi = \frac{\sqrt{\bar{n}}}{2}\int_t^{t'} d\tau f(\tau) \xi^\dagger(\tau) \;\text{ for } \xi \in\{ c,d \} \label{eq:Zop}
\end{align}

As evident from Eq. \ref{eq:Lambda_TQE}, a detection event in the bright port guarantees that at least one of the photons was extracted at some earlier time but it does not provide information on the time ordering between the two extracted photons. In the dark port, detection events can only occur within the timeframe between the two extractions. Since $\bigl[e^{Z_d} + (-1)^m e^{-Z_d}\bigr]$ for $m=0$ ($m=1$) comprises only of even (odd) powers of creation operator $Z_d$, the parity of detection events in the dark port collapses the outgoing state of the extracted photons into two possibilities; an even (odd) number of detection events heralds a temporal symmetric (antisymmetric) two-photon state. Alternatively, the extracted photons can be utilized to herald the generation of a cat state; bunching (antibunching) observed in a HOM experiment between the extracted photons indicates the preparation of an even (odd) cat state in the dark port.

To summarize, employing a TQE in the transmitted fields of two $\Lambda$-type sources allows us to herald the symmetry of the extracted photons, thereby recovering perfect two-photon interference from partially distinguishable photons. Unlike NL sources, since extraction is practically deterministic for $\bar{n}\! \gg\! 1$, the introduction of the TQE does not reduce the success probability for photon generation. Moreover, in an ideal setup, only a single photon is extracted from each source, avoiding number impurity concerns. While the deterministic nature of this source might suggest that generalizing the TQE scheme to more than two photons is straightforward, this task is not trivial, as we will describe in the Discussion section. Additionally, it should be noted that the presence of photon loss in realistic systems can hinder an accurate classification of the symmetry of the two-photon state, as the parity of detection events at the dark port is sensitive to such losses.

\section{Fusion gates}
\label{sec:Fusion}
The TQE enables the retrieval of ideal two-photon interference from a pair of imperfect SPS by decomposing the two-photon wavefunction into its symmetric and antisymmetric components. In turn, this allows us to reliably perform fusion gates with such sources, where our knowledge of the symmetry of the two-photon state is utilized to correctly interpret the destructive measurement outcome.

\begin{figure}
    \centering
    \includegraphics[width=\linewidth]{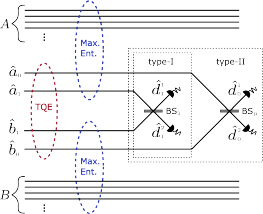}
    \caption{Fusion gates. Schematic description of type-I and type-II fusion gates between two path-encoded photonic qubits, defined in modes $\hat{a}_{0,1}$ and $\hat{b}_{0,1}$, each maximally entangled with a distinct set of qubits, $A$ and $B$, respectively. A type-I fusion gate is implemented by mixing modes $\hat{a}_{1}$ and $\hat{b}_{1}$ on a BS and registering photon detections at its output ports $\hat{d}_1^{1,2}$. The success of the gate is heralded by a single detection event in one of these ports, which results in a maximally entangled state of a new qubit, defined in $\hat{a}_{0}$ and $\hat{b}_{0}$, and qubit sets $A$ and $B$. In a type-II fusion, modes $\hat{a}_{0}$ and $\hat{b}_{0}$ are also mixed on a BS and the detection pattern in all output ports is recorded. A successful type-II fusion is heralded by two detection events in the output ports of different BS, which lead to a maximally entangled state of qubit sets $A$ and $B$. The symmetry of the two-photon state in modes $\hat{a}_{0,1}$ and $\hat{b}_{0,1}$, heralded by a TQE in the generation process of the photons, is used to correctly interpret the outcomes of these fusion gates (see Table \ref{table:1}).}
    \label{fig:Fusion}
\end{figure}

As depicted in Figure \ref{fig:Fusion}, consider two path-encoded photonic qubits, defined in modes $\hat{a}_{0,1}$ and $\hat{b}_{0,1}$, where each qubit is maximally entangled with a distinct set of qubits, $A$ and $B$, respectively. When the single photon in qubit modes $\hat{a}_{0,1}$ ($\hat{b}_{0,1}$) originates from the same source, the state can be written as,
\begin{align}
    \ket{\psi} \propto \int dt_1dt_2 &\phi(t_1,t_2) \Bigl(\ket{A_0}a_0^\dagger(t_1) + \ket{A_1}a_1^\dagger(t_1)\Bigr) \\
    &\times \Bigl(\ket{B_0}b_0^\dagger(t_2) + \ket{B_1}b_1^\dagger(t_2)\Bigr) \ket{0} \nonumber 
\end{align}
For pure single photons in modes $\hat{a}_{0,1}$ and $\hat{b}_{0,1}$, the temporal wavefunction is separable and can be expressed as $\phi(t_1,t_2) =  f(t_1)g(t_2)$, where $f$ and $g$ represent the temporal mode of $\hat{a}$ and $\hat{b}$, respectively. However, when applying the TQE to two modally impure photon sources, the two-photon state is projected into its temporally symmetric or antisymmetric part, resulting in a joint temporal wavefunction $\phi(t_1,t_2) =\varphi_m(t_1,t_2)$ that is either symmetric ($m=0$) or antisymmetric ($m=1$), maintaining,
\begin{align}
    \varphi_m(t_1,t_2)  = (-1)^m \varphi_m(t_2,t_1) 
\end{align}
It is important to crucial that the temporal symmetry of the two-photon state does not impose any constraints on the state of the path-encoded qubits, as they relate to different photonic degrees of freedom.

A type-I fusion gate involves mixing modes $\hat{a}_1$ and $\hat{b}_1$ on a BS and placing a detector at each output port. In a type-II fusion gate, modes $\hat{a}_0$ and $\hat{b}_0$ are also mixed on a  BS and the detection pattern of all four detectors is recorded. We denote the detection modes by $\hat{d}^j_k$, where $k\! \in\! \{0,1\}$ represents the BS mixing modes $\hat{a}_k$ and $\hat{b}_k$, and $j \!\in\! \{ 1,2\}$ specifies the two output ports of that BS.

\begin{table*}[t]
\centering 
\caption{Fusion gates outcomes of the scheme in Figure \ref{fig:Fusion}. A single detection event recorded in either output of $\text{BS}_1$ indicates a successful type-I fusion. This gate results in a maximally entangled state between qubit sets $A$ and $B$, along with a new qubit whose states $\ket{0}$ and $\ket{1}$ are defined by a single photon in modes $\hat{a}_0$ and $\hat{b}_0$, respectively. A successful type-II fusion gate involves two detection events at the output ports of different BS, resulting in a maximally entangled state for qubit sets $A$ and $B$. The successful outcomes in both fusion gates depend on the location of the detection events as well as on the symmetry of the incident two-photon state, as indicated by the sign of the superposition states. Failure of the fusion gates, leading to a separable state for qubit sets $A$ and $B$, occurs when two detection events are recorded at the output ports of the same BS.}
\begin{tabular}{ c c c c }
\hline \hline
Fusion & Detection events & Symmetric (m=0) & Antisymmetric (m=1)\\
\hline
Type-I & one event: $d^j_1$ & $\ket{A_0,B_1,0}\!+\!(-1)^j\ket{A_1,B_0,1}$ & $\ket{A_0,B_1,0}\!-\!(-1)^j\ket{A_1,B_0,1}$ \\
Type-II & two events: $d^i_0 d^j_1$ & $\ket{A_0,B_1}\!+\!(-1)^{i+j}\ket{A_1,B_0}$ & $\ket{A_0,B_1}\!-\!(-1)^{i+j}\ket{A_1,B_0}$ \\
Failure & two events: $d^i_{k}d^j_{k}$ & $\ket{A_{k}}\ket{B_{k}}$ ($i \!= \!j$) & $\ket{A_{k}}\ket{B_{k}}$ ($i\! \ne \! j$) \\
\hline\hline
\end{tabular} 
\label{table:1}
\end{table*}

A successful type-I fusion gate is heralded by a single detection event in either $\hat{d}^1_{1}$ or $\hat{d}^2_{1}$. As shown by Eq. \ref{eq:fusion_typeI}, qubit sets A and B, along with a new photonic qubit defined in modes $\hat{a}_0$ and $\hat{b}_0$, assume a maximally entangled state with a phase that depends on the symmetry of the two-photon state and the location of the detection event.
\begin{alignat}{2}
    d^j_{1}(t) \ket{\psi_m} \propto   \label{eq:fusion_typeI}  &\int &&dt' \varphi_m(t,t') \\
    &\times &&\Bigl( \ket{A_1}\ket{B_0} b^\dagger_0(t') \nonumber \\ & &&+(-1)^{m+j} \ket{A_0}\ket{B_1} a^\dagger_0(t') \Bigr)\ket{0} \nonumber
\end{alignat}

The success of a type-II fusion gate is signaled by a pair of detection events, each at the output of a different BS. As indicated by Eq. \ref{eq:fusion_typeII}, the final state of qubit sets $A$ and $B$ is maximally entangled, where the relative phase in the superposition is determined by the detection pattern and the exchange symmetry of the incident two-photon state.
\begin{alignat}{2}
    d^i_{0}(t)d^j_{1}(t') \ket{\psi_m} \propto &    
    (-1)^j \varphi_m(t,t')  \label{eq:fusion_typeII}\\
    &\times\Bigl(\ket{A_0}\ket{B_1} + (-1)^{m+i+j} \ket{A_1}\ket{B_0} \Bigr) \nonumber
\end{alignat}

In contrast, as evident in Eq. \ref{eq:fusion_typeII_fail}, two detection events at the outputs of the same BS collapse the state of qubit sets $A$ and $B$ to a product state. This results in a failure of the fusion, either type-I when $k\!=\!1$ or type-II for any value of $k$.
\begin{align}
    d^i_{k}(t)d^j_{k}(t') \ket{\psi_m} \propto   \label{eq:fusion_typeII_fail}  
    &(-1)^j \varphi_m(t,t') \\
    &\times \Bigl( 1 + (-1)^{m+i+j} \Bigr)\ket{A_k}\ket{B_k} \nonumber
\end{align}

Overall, the different outcomes of the fusion gates (up to normalization and a global sign) are summarized in Table \ref{table:1}. While the success probability of these gates remains the well-known $1/2$, employing a TQE and heralding on the symmetry of the two-photon state eliminates the infidelity arising from photon distinguishability. In comparison, implementing a fusion gate with photons that are not completely indistinguishable, i.e. their two-photon wavefunction is not purely symmetric or antisymmetric, results in mixing output states with different phases, given by the value of $m$ in Eq. \ref{eq:fusion_typeI} and \ref{eq:fusion_typeII}, which leads to a lower degree of entanglement and, hence, gate infidelity. 

\section{Discussion}
\label{sec:Diss}
We have shown that two-photon linear-optics gates rely on the exchange symmetry of the input two-photon wavefunction, and that identical photons (which indeed fall into the category of a symmetric wavefunction) are only a subset of the useful photonic states. Accordingly, even with sources that inherently produce photons in distinct modes due to temporal information leakage during the generation process, perfect two-photon interference can still be achieved from pairs of such sources. This is accomplished by applying a TQE on this information, leading to the heralded symmetrization or antisymmetrization of the spatial part of the two-photon wavefunction, accompanied by a symmetric or antisymmetric joint temporal profile, respectively.
We note that this application of the quantum eraser acts on a multimode continuous degree of freedom and not on a discrete degree of freedom, which is typically the case, such as polarization \cite{kwiat1992observation} or time-bins \cite{rudolph2023photonic}. As both TQE outcomes exhibit unit visibility in two-photon interference, this approach facilitates a high fidelity implementation of two-photon linear-optics gates, such as fusion gates, with SPS emitting partially distinguishable photons. Furthermore, in $\Lambda$-type sources, the TQE can be used in reverse to deterministically generate cat states.

The main drawback of implementing the TQE scheme between a pair of NL sources is the introduction of number impurity to the output state, caused by the generation of double photon pairs from a single source. This limitation restricts the use of the two-photon state in protocols that require no more than a single photon per rail. However, two-photon states with a well-defined temporal symmetry, even with multiple photons in each rail, may still hold value. Further research is needed to explore whether such states can be utilized in linear-optics operations that involve more than one photon per rail, such as \cite{ewert20143,gimeno2015three}.

On the other hand, in $\Lambda$-type sources, the primary challenge is the susceptibility to photon loss and imperfect single-photon detection efficiency, which can mislead the parity measurement of the number of photons at the dark port of the interferometer. Moreover, in realistic quantum-emitter systems, the state of the photonic field following single-photon extraction may deviate from the ideal form described in Eq. \ref{eq:extract} due to system-specific imperfections. As a result, the practical applicability of the TQE scheme is highly dependent on the details of the physical system in use. Nonetheless, drastically reducing loss and improving photon detection efficiency are major efforts in the field of photonic quantum computation. These efforts include, for example, the integration of highly efficient superconducting nanowire single-photon detectors \cite{marsili2013detecting,reddy2020superconducting,alexander2024manufacturable}. In parallel, efficient photon-emitter interfaces capable of performing high-fidelity operations at the single-photon level are being developed across various platforms \cite{tiecke2014nanophotonic,scheucher2016quantum,bechler2018passive,brekenfeld2020quantum,uppu2021quantumdotbased,schupp2021interface,knall2022efficient,reiserer2022cavity}. Specifically, the feasibility of the TQE scheme has been investigated in a system of single atoms coupled to optical resonators, accounting for imperfect extraction operations in the presence of photon loss channels \cite{aqua2024complete}.

The proposed scheme focuses on symmetrization of the two-photon wavefunction generated by a pair of SPS. Generalizing this method to more than a pair of photons falls beyond the scope of this paper; while a general two-photon wavefunction can be straightforwardly decomposed into its symmetric and antisymmetric part by adding or subtracting the two permutations, the process becomes more complex for larger multiphoton states, involving additional permutation eigenstates. Nevertheless, it may be possible to construct a more intricate, and potentially probabilistic, TQE tailored to meet the specific requirements of linear-optics operations involving more than two photons.

\begin{acknowledgments}
We acknowledge support from the Israeli Science Foundation, the Binational Science Foundation, H2020 Excellent Science (DAALI 899275), and the Minerva Foundation. B.D. is the Dan Lebas and Roth Sonnewend Professorial Chair of Physics.
\end{acknowledgments}

\begin{widetext}

\appendix
\section{TQE with multi-pair photon production in NL sources}
\label{app:NL_multipair}
In the TQE scheme described in Section \ref{sec:NL}, we focus on the scenario where each NL source produces a single pair of signal and idler photons. However, two detection events in $\hat{c}$ and $\hat{d}$ (see Figure \ref{fig:NL}) are just as likely to result from either one of the sources generating two pairs of signal and idler photons. We examine the operation of the TQE in this case. The relative phase of the pump between the two sources can be adjusted such that the state at the output of the NL media reads,
\begin{alignat}{1}
     \ket{\varphi^{(2)}_\text{NL}} = \frac{1}{2} \int dt_1 dt_2 dt_3 dt_4 \Phi(t_1,t_2) \Phi(t_3,t_4) \Bigl(&a_i^\dagger(t_1)a_s^\dagger(t_2)a_i^\dagger(t_3)a_s^\dagger(t_4) \\
     & + b_i^\dagger(t_1)b_s^\dagger(t_2)b_i^\dagger(t_3)b_s^\dagger(t_4)\Bigr)\ket{0} \nonumber
\end{alignat}
Following the BS operation,
\begin{alignat}{1}
     \ket{\varphi^{(2)}_\text{NL}} = \frac{1}{4}\int dt_1 dt_2 dt_3 dt_4 \Phi(t_1,t_2) \Phi(t_3,t_4)  \Biggl[&\Bigl(c^\dagger(t_1)c^\dagger(t_3) + d^\dagger(t_1)d^\dagger(t_3)\Bigr)   \Bigl(a_s^\dagger(t_2)a_s^\dagger(t_4) + b_s^\dagger(t_2)b_s^\dagger(t_4)\Bigr) \label{eq:NL2f} \\
     &+\Bigl(c^\dagger(t_1)d^\dagger(t_3) + c^\dagger(t_3)d^\dagger(t_1)\Bigr)  \Bigl(a_s^\dagger(t_2)a_s^\dagger(t_4) - b_s^\dagger(t_2)b_s^\dagger(t_4)\Bigr)\Biggr]\ket{0} \nonumber
\end{alignat}
As specified by Eq. \ref{eq:NL2f}, coalescent (coincident) detection events at the output ports of the BS herald a symmetric (antisymmetric) two-photon state, in a similar manner to the result of Eq. \ref{eq:NLf}. However, unlike the state in Eq. \ref{eq:NLf}, in this case, both photons occupy either mode $\hat{a}_s$ or $\hat{b}_s$. Interestingly, since the symmetry of the two-photon wavefunction dictates the outcome of a HOM experiment, perfect two-photon interference of the signal photons persists even when a double pair is generated by either one of the two NL sources. Specifically, the symmetric (antisymetric) state of the signal photons in \ref{eq:NL2f} is characterized by a coincidence rate of zero (one).

\section{TQE with SPRINT-based sources}
\label{app:SPRINT_TQE}
The state of the photonic field, resulting from the successful extraction of a single photon to mode $\hat{a}_r$ from an incident coherent pulse in $\hat{a}_l$, with an average photon number $\bar{n}$ and an temporal envelope $f(t)$, is given by \cite{gea2013photon},
\begin{alignat}{1}
    \ket{\psi^{(1)}_\Lambda} \propto &\int_{-\infty}^{\infty} dt f(t) a_r^\dagger(t) 
    \text{exp}\biggl(\sqrt{\bar{n}} \int_{t}^{\infty} dt' f(t') a_l^\dagger(t') \biggr) \ket{0} 
\end{alignat}
We consider two extraction sources in the configuration depicted in Figure \ref{fig:Lambda}(b), where a coherent pulse is split by a 50:50 BS and directed to each of the extraction sources. The state of the field can then be written as the tensor product of two individual sources, each  incident by an incoming coherent pulse with an average of $\frac{\bar{n}}{2}$ photons,
\begin{alignat}{1}
    \ket{\psi^{(1)}_\Lambda} \propto &\int_{-\infty}^\infty dt f(t) a_r^\dagger(t)
    \text{exp}\biggl(\sqrt{\frac{\bar{n}}{2}} \int_{t}^{\infty} dt' f(t') a_l^\dagger(t') \biggr) \\
    \otimes &\int_{-\infty}^\infty d\tau f(\tau) b_r^\dagger(\tau)
    \text{exp}\biggl(\sqrt{\frac{\bar{n}}{2}} \int_{\tau}^{\infty} d\tau' f(\tau') b_l^\dagger(\tau') \biggr)\ket{0} \nonumber
\end{alignat}
We apply the following BS transformation on the transmitted modes $\hat{a}_l$ and $\hat{b}_l$,
\begin{align}
    c = \frac{a_l+b_l}{\sqrt{2}} \;\;\; ; \;\;\; d = \frac{a_l-b_l}{\sqrt{2}}
\end{align}
and get the state,
\begin{alignat}{1}
    \ket{\psi^{(1)}_\Lambda} \propto &\int_{-\infty}^\infty dt f(t) a_r^\dagger(t) 
    \text{exp}\biggl(\frac{\sqrt{\bar{n}}}{2} \int_{t}^{\infty} dt' f(t') \bigl(c^\dagger(t') + d^\dagger(t')\bigr) \biggr)  \\
    \otimes &\int_{-\infty}^\infty d\tau f(\tau) b_r^\dagger(\tau) \nonumber
    \text{exp}\biggl(\frac{\sqrt{\bar{n}}}{2} \int_{\tau}^{\infty} d\tau' f(\tau') \bigl(c^\dagger(\tau')- d^\dagger(\tau') \bigr) \biggr)\ket{0} \nonumber
\end{alignat}
Rearranging the terms,
\begin{alignat}{1}
    \ket{\psi^{(1)}_\Lambda} \propto &\int_{-\infty}^\infty dt d\tau f(t)f(\tau) a_r^\dagger(t) b_r^\dagger(\tau) 
    \text{exp}\biggl(\frac{\sqrt{\bar{n}}}{2} \biggl[\int_{t}^{\infty} dt' f(t') c^\dagger(t') +  \int_{\tau}^{\infty} d\tau' f(\tau') c^\dagger(\tau') \biggr]\biggr) \label{eq:appB5}\\
    &\times \text{exp}\biggl(\frac{\sqrt{\bar{n}}}{2} \biggl[\int_{t}^{\infty} dt' f(t') d^\dagger(t') -  \int_{\tau}^{\infty} d\tau' f(\tau') d^\dagger(\tau') \biggr]\biggr) \ket{0} \nonumber
\end{alignat}
and using,
\begin{align}
    \int_a^b dx + \int_b^c dx = \int_a^c dx
\end{align}
we can write Eq. \ref{eq:appB5} as,
\begin{alignat}{1}
    \ket{\psi^{(1)}_\Lambda} \propto &\int_{-\infty}^\infty dt d\tau f(t)f(\tau) a_r^\dagger(t) b_r^\dagger(\tau)   
     \text{exp}\biggl(\sqrt{\bar{n}} \int_{\text{max}(t,\tau)}^{\infty} dt' f(t') c^\dagger(t') \biggr) 
    \text{exp}\biggl(\frac{\sqrt{\bar{n}}}{2}  \int_{\text{min}(t,\tau)}^{\text{max}(t,\tau)} d\tau' f(\tau') c^\dagger(\tau') \biggr) \\
    &\times \text{exp}\biggl(\frac{\sqrt{\bar{n}}}{2} \int_{t}^{\tau} dt' f(t') d^\dagger(t')\biggr) \ket{0} \nonumber
\end{alignat}
We decompose the double integral $dtd\tau$ using,
\begin{align}
    \int_{-\infty}^{\infty} dtd\tau = \int_{-\infty}^{\infty} dt \int_{t}^{\infty} d\tau + \int_{-\infty}^{\infty} dt \int_{-\infty}^{t} d\tau 
\end{align}
and get, 
\begin{alignat}{1}
    \ket{\psi^{(1)}_\Lambda} \propto \Biggl[ &\int_{-\infty}^\infty dt \int_{t}^\infty d\tau f(t)f(\tau) a_r^\dagger(t) b_r^\dagger(\tau)   
     \text{exp}\biggl(\sqrt{\bar{n}} \int_{\tau}^{\infty} dt' f(t') c^\dagger(t') \biggr) 
    \text{exp}\biggl(\frac{\sqrt{\bar{n}}}{2}  \int_{t}^{\tau} d\tau' f(\tau') c^\dagger(\tau') \biggr) \\
    &\times \text{exp}\biggl(\frac{\sqrt{\bar{n}}}{2} \int_{t}^{\tau} dt' f(t') d^\dagger(t')\biggr) \nonumber \\
    & + \int_{-\infty}^\infty dt \int_{-\infty}^t d\tau f(t)f(\tau) a_r^\dagger(t) b_r^\dagger(\tau)   
     \text{exp}\biggl(\sqrt{\bar{n}} \int_{t}^{\infty} dt' f(t') c^\dagger(t') \biggr) 
    \text{exp}\biggl(\frac{\sqrt{\bar{n}}}{2}  \int_{\tau}^{t} d\tau' f(\tau') c^\dagger(\tau') \biggr) \nonumber \\
    &\times \text{exp}\biggl(\frac{\sqrt{\bar{n}}}{2} \int_{t}^{\tau} dt' f(t') d^\dagger(t')\biggr) \Biggr] \ket{0} \nonumber
\end{alignat}
Renaming the integration variables $t\leftrightarrow\tau$ in the second double integral,
\begin{alignat}{1}
    \ket{\psi^{(1)}_\Lambda} \propto \Biggl[ &\int_{-\infty}^\infty dt \int_{t}^\infty d\tau f(t)f(\tau) a_r^\dagger(t) b_r^\dagger(\tau)   
     \text{exp}\biggl(\sqrt{\bar{n}} \int_{\tau}^{\infty} dt' f(t') c^\dagger(t') \biggr) 
    \text{exp}\biggl(\frac{\sqrt{\bar{n}}}{2}  \int_{t}^{\tau} d\tau' f(\tau') c^\dagger(\tau') \biggr)  \\
    &\times \text{exp}\biggl(\frac{\sqrt{\bar{n}}}{2} \int_{t}^{\tau} dt' f(t') d^\dagger(t')\biggr) \nonumber \\
    & + \int_{-\infty}^\infty d\tau \int_{-\infty}^\tau dt f(t)f(\tau) a_r^\dagger(\tau) b_r^\dagger(t)   
     \text{exp}\biggl(\sqrt{\bar{n}} \int_{\tau}^{\infty} dt' f(t') c^\dagger(t') \biggr) 
    \text{exp}\biggl(\frac{\sqrt{\bar{n}}}{2}  \int_{t}^{\tau} d\tau' f(\tau') c^\dagger(\tau') \biggr) \nonumber \\
    &\times \text{exp}\biggl(\frac{\sqrt{\bar{n}}}{2} \int_{\tau}^{t} dt' f(t') d^\dagger(t')\biggr) \Biggr] \ket{0} \nonumber
\end{alignat}
and since,
\begin{align}
    \int_{-\infty}^{\infty} d\tau \int_{-\infty}^{\tau} dt = \int_{-\infty}^{\infty} dt \int_{t}^{\infty} d\tau 
\end{align}
we can write,
\begin{alignat}{1}
    \ket{\psi^{(1)}_\Lambda} \propto &\int_{-\infty}^\infty dt \int_{t}^\infty d\tau f(t)f(\tau)    
     \text{exp}\biggl(\sqrt{\bar{n}} \int_{\tau}^{\infty} dt' f(t') c^\dagger(t') \biggr) 
    \text{exp}\biggl(\frac{\sqrt{\bar{n}}}{2}  \int_{t}^{\tau} d\tau' f(\tau') c^\dagger(\tau') \biggr)  \\
    &\times \biggl[ a_r^\dagger(t) b_r^\dagger(\tau)\text{exp}\biggl(\frac{\sqrt{\bar{n}}}{2} \int_{t}^{\tau} dt' f(t') d^\dagger(t')\biggr)  + a_r^\dagger(\tau) b_r^\dagger(t)\text{exp}\biggl(-\frac{\sqrt{\bar{n}}}{2} \int_{t}^{\tau} dt' f(t') d^\dagger(t')\biggr) \biggl] \ket{0} \nonumber
\end{alignat}
Using the notation in Eq. \ref{eq:Zop},
\begin{align}
    Z_\xi = \frac{\sqrt{\bar{n}}}{2}\int_t^{\tau} dt' f(t') \xi^\dagger(t') \;\text{ for } \xi \in\{ c,d \}
\end{align}
we can write the state as,
\begin{alignat}{1}
    \ket{\psi^{(1)}_\Lambda} \propto \int_{-\infty}^\infty dt \int_{t}^\infty d\tau f(t)f(\tau)    
    \text{exp}\biggl(\sqrt{\bar{n}} \int_{\tau}^{\infty} dt' f(t') c^\dagger(t') \biggr) 
    e^{Z_c} 
     \biggl[ a_r^\dagger(t) b_r^\dagger(\tau)e^{Z_d}  + a_r^\dagger(\tau) b_r^\dagger(t)e^{-Z_d} \biggl] \ket{0} \nonumber \\
\end{alignat}
Finally, we rearrange the terms in square brackets using the following identity,
\begin{align}
    AB+CD=\frac{1}{2}[(A+C)(B+D) + (A-C)(B-D)]
\end{align}
and obtain the state in Eq. \ref{eq:Lambda_TQE},
\begin{alignat}{1}
    \ket{\psi^{(1)}_\Lambda} \propto \int_{-\infty}^\infty dt \int_{t}^\infty d\tau & f(t)f(\tau)    
    \text{exp}\biggl(\sqrt{\bar{n}} \int_{\tau}^{\infty} dt' f(t') c^\dagger(t') \biggr) 
    e^{Z_c}  \\
     \times\frac{1}{2}\biggl[& (a_r^\dagger(t) b_r^\dagger(\tau) +  a_r^\dagger(t) b_r^\dagger(\tau))(e^{Z_d}+e^{-Z_d}) \nonumber\\
     &+ (a_r^\dagger(t) b_r^\dagger(\tau) -  a_r^\dagger(t) b_r^\dagger(\tau))(e^{Z_d}-e^{-Z_d}) \biggl] \ket{0} \nonumber 
\end{alignat}

\end{widetext}

\bibliography{refs.bib}

\end{document}